\documentclass[letterpaper,twocolumn,10pt]{article}
\usepackage{usenix-2020-09}
\usepackage{tikz}
\usepackage{xcolor,xspace}
\usepackage{url}
\usepackage{epsfig,graphicx,endnotes,kotex,subfig,multirow,amsmath,algorithm,algpseudocode}
\usepackage{balance}
\usepackage{enumitem}
\usepackage{flushend}
\usepackage[T1]{fontenc}
\usepackage{color,soul}
\usepackage{amsthm}

\begin{document}

\newcommand{\sys}{OrbitCache\xspace}
\newcommand{\sota}{NetCache\xspace}
\newcommand{\baseline}{NoCache\xspace}
\newcommand{\etal}{\textit{et al.}\xspace}


\title{\Large \bf Pushing the Limits of In-Network Caching for Key-Value Stores}
\date{}
\author{
{\rm Gyuyeong Kim }\\ Sungshin Women's University
}
\maketitle
\begin{abstract}
We present OrbitCache, a new in-network caching architecture that can cache variable-length items to balance a wide range of key-value workloads.
Unlike existing works, OrbitCache does not cache hot items in the switch memory.
Instead, we make hot items revisit the switch data plane continuously by exploiting packet recirculation.
Our approach keeps cached key-value pairs in the switch data plane while freeing them from item size limitations caused by hardware constraints.
We implement an OrbitCache prototype on an Intel Tofino switch.
Our experimental results show that OrbitCache can balance highly skewed workloads and is robust to various system conditions. 
\end{abstract}

\section{Introduction \label{introduction}}
Key-value stores are the fundamental building blocks for online services owing to fast access to variable-length data~\cite{redis,memcached,rocksdb,mica}.
Workloads are generally read-intensive, and most items are hundreds of bytes~\cite{cao20, atikoglu12, blott15}.
A challenge in scaling key-value stores is mitigating load imbalance among storage servers caused by skewed key popularity (e.g., trending events).
The load imbalance leads to overloading hot item servers, causing overall performance degradation~\cite{netcache,switchkv,pegasus}.

In-network caching is a promising load-balancing approach that builds a cache in the switch by leveraging programmable switch ASICs like Intel Tofino~\cite{tofino}.
The idea is to store hot items in the on-chip switch memory.
Requests read the item value by referring to the cache lookup table where the item key is the table index.
Compared to server-based caching, in-network caching provides much higher performance without extra cache nodes or accelerators thanks to a Tbps-scale processing throughput.

Unfortunately, existing solutions~\cite{netcache,farreach,distcache} can cache only tiny items due to hardware constraints.
Specifically, the key size is limited to 16 bytes due to the maximum match-key width of the match-action table.
A limited number of match-action stages and a small accessible byte size per stage make it difficult for the value size to exceed 128 bytes.
These numbers are far from typical workloads where key and value sizes are still small enough to be stored in a single packet, but larger than the limit~\cite{atikoglu12,nishtala13,cao20,blott15,yang20}.
For example, in 54 Twitter workloads~\cite{yang20}, most keys are tens of bytes and many values are less than 1024 bytes.
However, the existing solutions cannot cache even a single item in 42 out of 54 workloads because the keys and values are generally larger than the limits.
Between the other 12 workloads, only 2 workloads have a portion of cacheable items larger than 50\%.
This trend is similar in Facebook workloads as well~\cite{cao20}.
In this context, we ask the following question: \textit{how can we cache variable-length items in the programmable switch for balancing typical key-value workloads?}

We present \sys, a new in-network caching architecture that can cache variable-length key-value items in the switch data plane.
Our idea is to make hot items visit the switch data plane continuously in the form of cache packets instead of caching the hot items in the switch memory.
We efficiently use packet recirculation, a built-in feature of the programmable switch ASIC.
This enables a packet to revisit the switch data plane through an internal loopback port without going outside. 
The switch can cache variable-length key-value pairs without being limited by the hardware constraints owing to circulating cache packets.
For cache serving, the switch maintains small request metadata in the switch memory and forwards the cache packet to the client if there is a pending request for the key.
To support variable-length keys, we use key hashes for cache lookups and resolve hash collisions at the client by comparing the requested key and the returned key, which is contained in the cache packet.

The key insight behind the idea is that it is hard to overcome the memory access constraint of switch hardware if we stick to the approach of caching items in the switch memory.
For example, the idea of using key hashes for variable-length keys is difficult to realize in the existing solutions.
This is because the switch does not have enough hardware resources to store keys and values together.
In \sys, cache packets contain both keys and values, allowing to handle hash collisions at the client.
Meanwhile, we may recirculate requests multiple times to read larger values.
However, this limits scalability because the number of in-flight packets in the recirculation port increases proportionately to the number of requests, making a bottleneck.
In addition, the switch has one internal recirculation port.
\sys avoids a bottleneck in the recirculation port.
This is because 1) requests are never recirculated; 2) the number of in-flight cache packets is small and constant; 3) the switch can process multiple cache packets simultaneously with hardware-level parallelism, thus minimizing queueing between cache packets.

Realizing the idea imposes various technical challenges.
First, the switch should buffer multiple request metadata for cached items.
We design a request table as a circular queue structure using multiple register arrays. 
Our request table provides isolated data access among different keys.
Second, a cache packet should serve multiple requests once fetched, rather than a single request.
To achieve this, the switch clones the cache packet with low overhead before forwarding it to the client using the packet replication engine (PRE), a special module of the switch ASIC.
Next, we should ensure cache coherence.
We design an invalidation-based coherence protocol that can fetch new values and send back a reply to the client simultaneously.
Lastly, the switch should deal with dynamic workloads where key popularity changes over time.
We design an efficient cache update mechanism in the switch control plane.

We have implemented a \sys prototype on an Intel Tofino switch.
To evaluate \sys, we build a testbed consisting of commodity servers.
Our experimental results show that \sys provides load balancing for many skewed workloads having diverse item sizes.
In addition, \sys is robust to various workload conditions like key access distributions, write ratio, and the number of servers.
We also show that \sys can adapt to dynamic workloads where key popularity changes over time.

In summary, this work makes the following contributions.
\begin{itemize}[noitemsep]
\item{
We design \sys, an in-network caching architecture where the switch can cache variable-length hot items to balance a wide range of workloads for distributed key-value stores.
}
\item{
We propose several techniques to address technical challenges when realizing the idea of variable-length in-network caching.
}
\item{
We implement a \sys prototype on Intel Tofino switches and show the efficiency and robustness of \sys through extensive testbed experiments.
}
\end{itemize}
\section{Background and Motivation \label{motivation}}
\subsection{In-Network Load-Balancing Caches}
\textbf{Balancing key-value stores with a small cache.}
In distributed key-value stores, balancing imbalanced loads across storage servers caused by different key popularity is a primary challenge.
Caching is a powerful technique to address the challenge, which is based on a theoretical result called the \textit{small cache effect}: we can balance loads for $N$ servers (or partitions) by caching the $O(N\log N$) hottest items, regardless of the number of items~\cite{fan11}.
Caching on commodity servers is a natural option to build the cache~\cite{memcached,fan11,switchkv}.
However, the performance of a cache node is not sufficient to handle many requests due to the limited throughput of CPUs.
Building a high-performance caching layer using multiple replicated cache nodes is expensive and causes poor write performance for cache coherence.

\begin{figure*}[t]
\centering\hfill
\subfloat[The \sota architecutre~\cite{netcache}]{\includegraphics[width=0.47\linewidth]{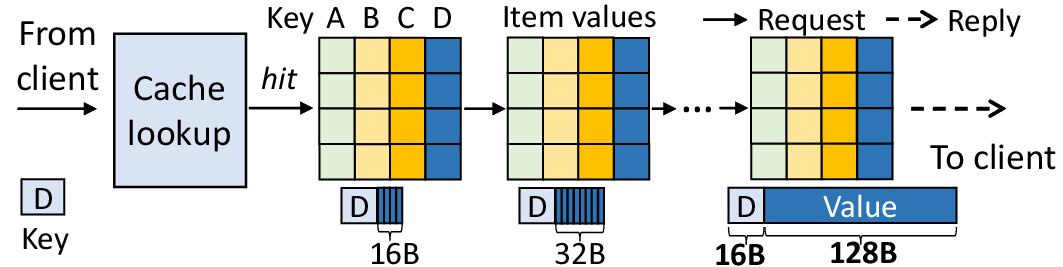}}\hfill
\subfloat[The proposed \sys architecture]{\includegraphics[width=0.47\linewidth]{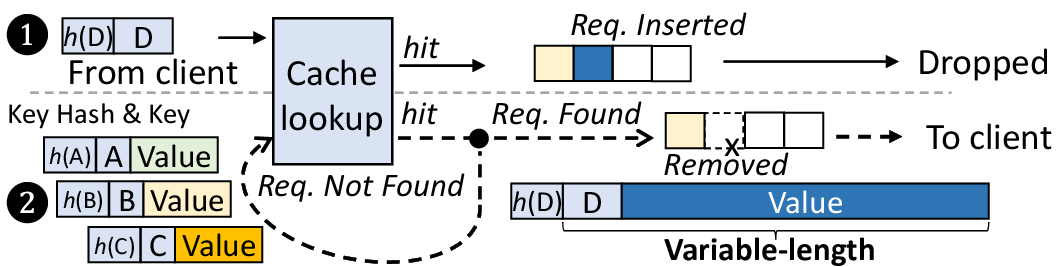}}\hfill
\caption{Comparison of the high-level idea with the \sota architecture.
\textit{
In \sys, clients submit requests, and then circulating cache packets read request metadata.
Since both keys and values are in cache packets, hardware constraints do not limit the item size.
For variable-length keys, \sys uses key hashes while handling hash collisions at the client. 
}
\label{fig:motiv-com}}
\end{figure*}

\textbf{Why in-network caching?}
Building a load-balancing cache in the switch by leveraging programmable switch ASICs like Intel Tofino~\cite{tofino} is an attractive approach.
Unlike server-based caching, this approach can provide high-performance load balancing in a single box without additional nodes.
This is because the switch is highly optimized for packet I/O and can process a few billion packets per second, which is an order of magnitude higher than that of the CPU in servers.
The switch also provides a low packet processing delay within hundreds of nanoseconds.

\textbf{Limited cacheable item size.}
Existing works~\cite{netcache,distcache,farreach} demonstrate the efficiency of in-network caching, but they limit the cacheable item size.
Specifically, they support items of up to 16-byte keys and 128-byte values.
In the reconfigurable match table (RMT) switch architecture~\cite{bosshart13} like Intel Tofino, the switch data plane consists of $n$ match-action stages~\cite{tofino}.
Each stage has a static memory and a few ALUs that can perform simple arithmetic operations on $k$ bytes.
The existing works store the value of cached items across multiple stages after fragmentation, limiting the maximum value size to $n\times k$ bytes.
Unfortunately, $n\times k$ of the current switch is quite small, and the available number of stages for accessing the value is less than $n$ since other non-caching functions also consume match-action stages.
The key size is also limited by the maximum match-key width of the match-action table in a single match-action stage.
This is because the existing works use a match-action table to implement the cache lookup table where the item key is the match key.

\textbf{Why is it not enough?}
Many key-value items are indeed small, but typically exceed the existing size limits. 
We have analyzed 54 Twitter workloads~\cite{yang20}, and observe that existing solutions are not enough to deal with typical workloads.
For example, only 3.7\% of the workloads have over 80\% of keys $\leq 16$ B.
38.9\% of the workloads have over 80\% of values $\leq 128$ B.
The existing works can cache less than 10\% of items for 85\% of the workloads because, to be cacheable, both key and value sizes must be within the limits.
Furthermore, they cannot cache even a single item for 77.8\% of the workloads.
Facebook workloads~\cite{cao20} show similar size distributions.
The average key size is 27.1 bytes, and the median value size is 235 bytes.
These numbers exceed the size limits of the existing solutions.

\subsection{Variable-Length In-Network Caching}
\textbf{Design rationale.}
We argue that it is hard to overcome the memory access constraint if we stick to the approach of caching data in the switch memory because the constraint is determined at the time of manufacture.
Therefore, we propose \sys, a different approach to in-network caching.

Our high-level idea is to keep cached key-value pairs circulating within the switch data plane through packet recirculation, a built-in feature of the programmable switch that makes the packet revisit the switch data plane.
In our approach, cached items read requests rather than the requests read the cached items.
Specifically, clients submit requests, and the switch maintains small request metadata like the client IP address.
\textit{Cache packets}, which are reply packets containing the key and the value of cached items, continue to revisit the data plane via a recirculation port while actively checking pending requests.
Note that cache packets do not go outside the switch.
If a request for the key is found, the matched cache packet is forwarded to the client.
Otherwise, the cache packet is recirculated.
In Figure~\ref{fig:motiv-com}, we illustrate the difference between \sys and the \sota architecture~\cite{netcache}, which is the reference architecture of existing in-network caching solutions~\cite{netcache,distcache,farreach}.

\textbf{Variable-length keys.}
Our approach enables us to support variable-length key-value pairs.
For variable-length keys, one possible solution within the limited match-key width of the match-action table is to use a fixed-size key hash as the match key.
To handle potential hash collisions, we should compare the requested key and the key of the value that has been read.
To do this, the switch data plane should also store the item keys alongside values.
Unfortunately, it is hard to realize in the existing architecture because there are not enough match-action stages to store both keys and values.
\sys can realize the idea of using key hashes since the switch maintains cached key-value pairs in the form of circulating reply packets.
Therefore, the client can get the correct value from the storage server if the requested and returned keys differ.

\textbf{Variable-length values.}
For variable-length values, a possible approach in the existing architecture is recirculating requests.
Specifically, requests can access the switch memory multiple times using packet recirculation.
However, this approach is not scalable because the number of in-flight packets per second in the recirculation port increases in proportion to the number of requests.
For example, if every request is recirculated 8 times to read a 1024-byte value, the effective throughput of the recirculation port is reduced to 1/8 of the bandwidth.
Unfortunately, a pipeline in the programmable switch has only one internal recirculation port, while there are tens of regular front ports.
This means that the recirculation port limits the performance excessively.

Our recirculation-based caching design avoids the potential bottleneck in the recirculation port as follows.
First, the switch never recirculates requests.
Second, only a small, constant number of cache packets are recirculated.
The time to recirculate and process a cache packet is a few hundred nanoseconds like normal packets.
Third, the switch can process multiple packets simultaneously through inter-stage parallelization, reducing queueing delay between cache packets.

\textbf{Trade-off.}
There is a trade-off of using recirculation to achieve variable-length caching.
Although we are free from size limitations, we should limit the cache size.
This is because, in \sys, requests should wait until cache packets handle them.
For a cache packet, the time to read a request is impacted by the other in-flight cache packets.
This means that the number of in-flight cache packets in the recirculation port determines the latency for cache serving.
Although the switch can process many cache packets simultaneously, the request may wait excessively until being served if there are too many cache packets.
The sacrifice of cache size for variable-length caching is backed by the small cache effect.
As described in Section 2.1, caching a small number of hot items is enough to balance skewed workloads~\cite{fan11,pegasus}.

\textbf{Technical challenges.}
To balance the trade-off and translate the idea into a working system, we should address several technical challenges as follows.
\begin{itemize}[noitemsep]
\item{
The switch should maintain multiple requests, especially for the same cached item.
If not, many requests for cached items would be forwarded to the server as they cannot be stored in the data plane.
}
\item{
A cache packet should serve multiple requests once fetched.
Otherwise, the switch must fetch the cache packet from the server again, degrading performance.
}
\item{
We should design a cache coherence mechanism between the switch and storage servers.
If not, a request may read a stale value for the requested key.
}
\item{
The switch should adapt to key popularity changes.
}
\end{itemize}
 
\begin{figure}[t!]
\centering
\includegraphics[width=8.0cm]{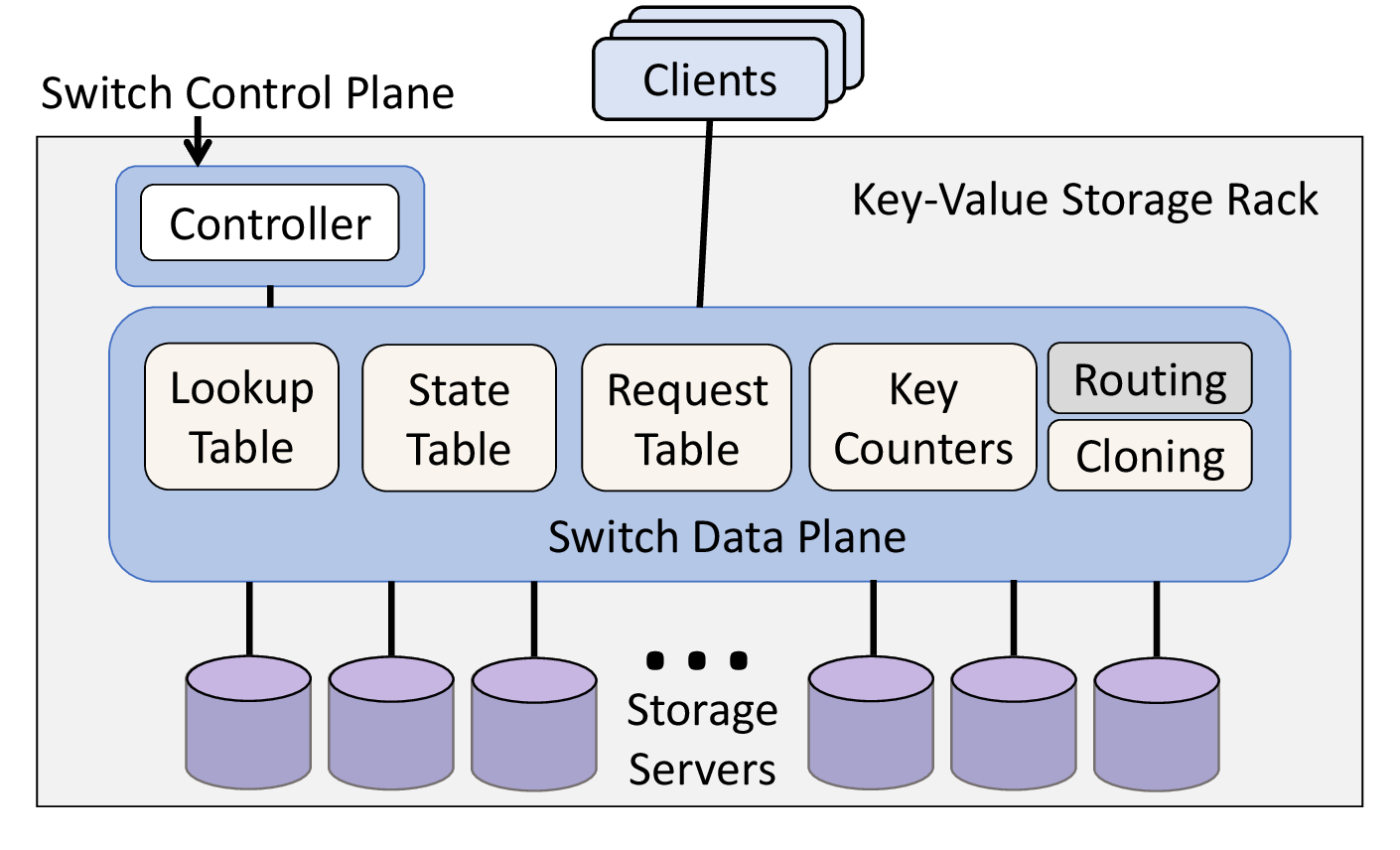}
\caption{\sys architecture. \label{fig:overview}}
\end{figure}
\section{\sys Design \label{design}}
\subsection{\sys Architecture}
Figure~\ref{fig:overview} shows the overview of the \sys architecture.

\textbf{Switch data plane.}
The switch data plane in \sys consists of several custom tables and modules as follows.
\begin{itemize}[noitemsep]
\item{
\textit{Lookup table} is for cache lookups, which is a match-action table that uses a hash of item keys as the match key.
The table returns a table index needed to access other tables and modules.
Cache entries are managed by the controller in the switch control plane.
}
\item{
\textit{State table}, implemented as a register array, maintains the validity of the value of cached items.
This is needed to prevent read requests from obtaining the stale value when there are pending writes for the requested key.
}
\item{
\textit{Request table}, implemented with multiple register arrays and registers, stores request metadata like client IP address, L4 port number, and request sequence number.
}
\item{
\textit{Key counters} consist of two registers and one register array.
The key popularity counter is a register array that tracks the key popularity for each key.
The cache hit counter and the overflow request counter are registers that track the total number of cache hits and the total number of overflow requests for all cached keys.
The controller uses these for cache sizing.
}
\item{
\textit{Cloning module} comprises a few match-action tables for cloning reply packets.
Packet cloning is done by the PRE, a hardware module in the switch ASIC~\cite{tofino}.
}
\end{itemize}

Meanwhile, the switch invokes the custom processing logic only for \sys packets by referring to reserved L4 ports.
We use UDP to handle messages for better latency like existing works~\cite{pegasus,switchkv,netcache} while using TCP for top-$k$ item reports in cache updates.
For normal packets, the switch only applies the traditional packet forwarding logic.

\textbf{Switch control plane.}
The controller in the switch control plane performs cache updates and switch configurations.
It evicts the least popular keys and inserts new hot keys based on server-side periodic top-$k$ reports for uncached keys and the switch-side report for cached keys.
Note that while the controller handles key insertion, value fetching is done via the switch data plane.
It also handles switch configurations like the rule update for packet forwarding tables.

\textbf{Clients and servers.}
Clients should specify the operation type, the item key to request, and the key hash.
Storage servers run a server application that acts as a shim layer that translates \sys messages to API calls for key-value stores and vice versa.
The server returns replies for read and write requests like regular storage servers.
However, in cases of write requests for cached items, the server returns a write reply with the item value as well to fetch the latest value.
\begin{figure}[t!]
\centering
\includegraphics[width=8.0cm]{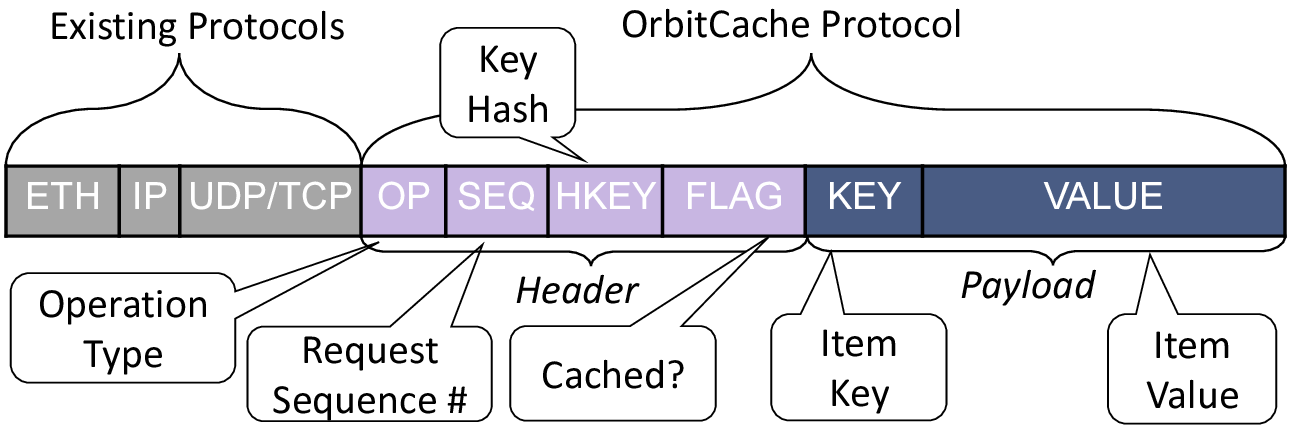}
\caption{\sys packet format.\label{fig:header}}
\end{figure}

\subsection{Packet Format}
Figure~\ref{fig:header} depicts the \sys packet format.
The message consists of the header and the payload.
The switch only parses the header.
The payload of \sys message consists of the key and the value.
Our header is 22 bytes.
Therefore, \sys supports a key-value pair of up to 1438 bytes for a single packet.
For example, the switch can cache an item with a 16-byte key and a 1422-byte value.
Our header fields are as follows.
\begin{figure*}[t]
\centering
\subfloat[Read request]{\includegraphics[width=0.49\linewidth]{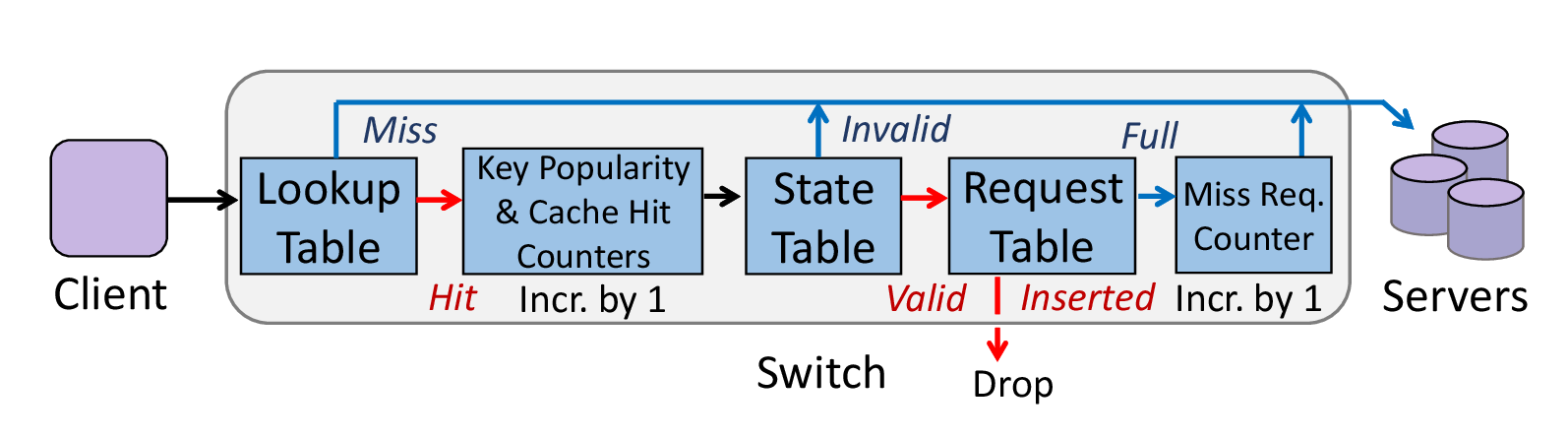}}\hfill
\subfloat[Read reply]{\includegraphics[width=0.49\linewidth]{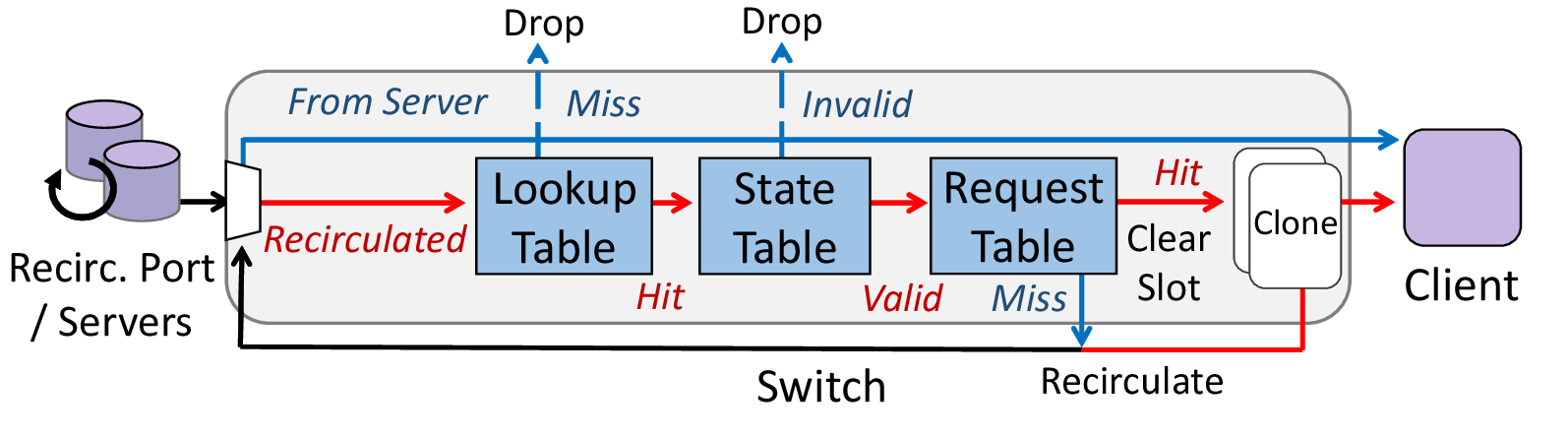}}\hfill
\subfloat[Write request]{\includegraphics[width=0.49\linewidth]{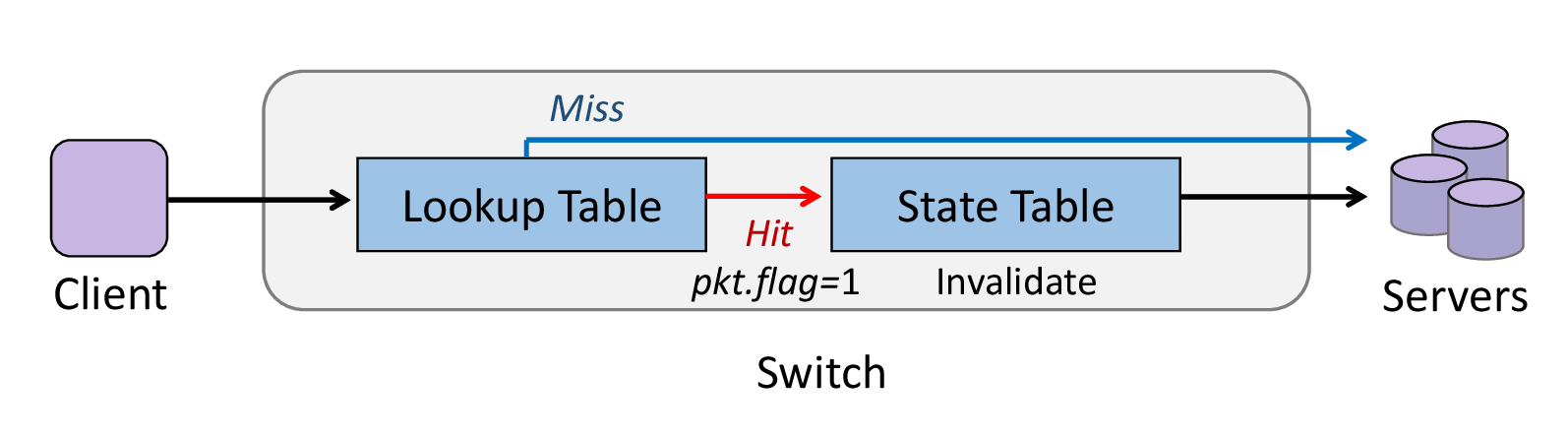}}\hfill
\subfloat[Write reply]{\includegraphics[width=0.49\linewidth]{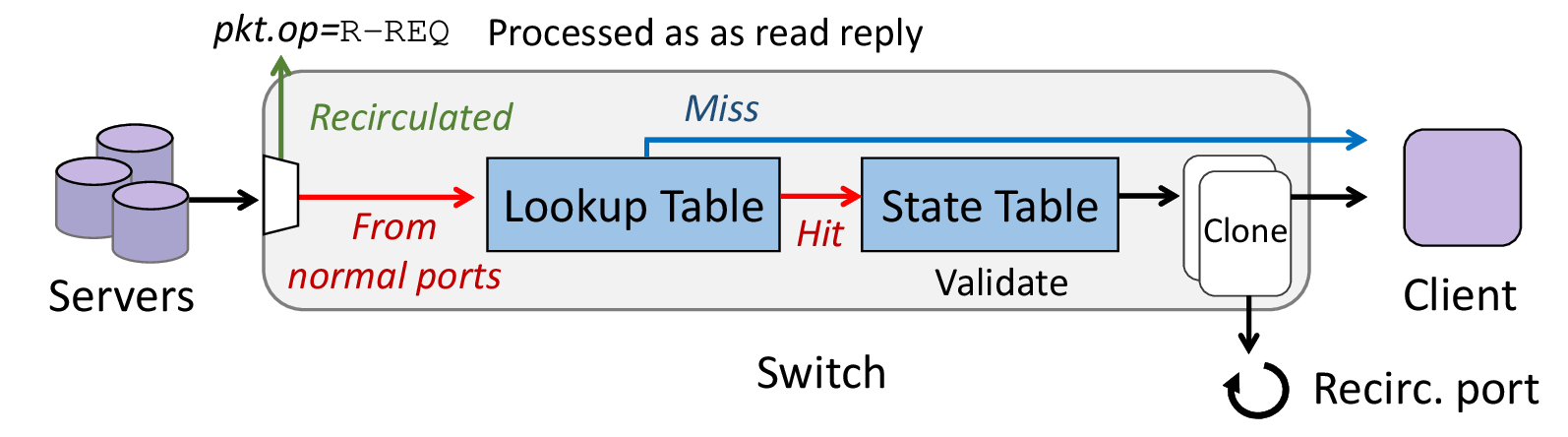}}\hfill
\caption{
Request processing.
\textit{
(a) the switch drops the request after inserting request metadata into the request table;
(b) If a circulating cache packet reads request metadata, the switch clones the packet so that the original packet is forwarded to the client and the cloned one is recirculated again for further serving;
(c) the switch invalidates the item to avoid inconsistent reads if a write request is for a cached item;
(d) upon receiving a write reply for a cached item, the switch validates the item. After that, the switch clones the packet.
The cloned packet is processed as a read reply after updating the operation type.
}
\label{fig:pro}}
\end{figure*}
\begin{itemize}[noitemsep]
\item{
\texttt{OP} (1 byte): the operation type, which can be \texttt{R-REQ} (Read request), \texttt{W-REQ} (Write request), \texttt{R-REP} (Read reply), \texttt{W-REP} (Write reply), \texttt{F-REQ} (Fetch request), \texttt{F-REP} (Fetch reply), and \texttt{CRN-REQ} (Correction request).
}
\item{
\texttt{SEQ} (4 bytes): a sequence number as request IDs for handling hash collisions.
}
\item{
\texttt{HKEY} (16 bytes): the key hash as cache lookup index.
}
\item{
\texttt{FLAG} (1 byte): a flag field to distinguish write requests for cached items from those for uncached items.
}
\end{itemize}

\subsection{Basic Request and Reply Processing}
In Figure~\ref{fig:pro}, we illustrate the packet processing logic.

\textbf{Request generation at clients.}
Clients should specify the \texttt{OP} and \texttt{HKEY} fields.
The \texttt{SEQ} field is increased by one for every request, which is used to resolve hash collisions.
The destination storage server is determined by hashing the key.

\textbf{Read requests.}
The switch first refers to the lookup table using a key hash to get a table index, which is used to access the slots in other tables.
If missed, the switch forwards the packet to the server as the request is for an uncached item.
In case of a cache hit, the key popularity counter and the cache hit counter are incremented by one.
This key popularity is collected by the controller in the switch control plane periodically for cache updates.
Next, the switch checks the validity of the requested key by looking into the state table.
The state is binary: valid or invalid.
Being invalid means that there are pending write requests for the key.
In this case, the switch forwards the read request to the server to avoid reading stale item values.
If the key is valid, the switch checks the request table to see whether there is a free slot.
The switch puts the request metadata into the table when a free slot is found.
Otherwise, the request is destined to the server after the overflow request counter is increased.
Request metadata includes the client IP address, L4 port number, and \texttt{SEQ} as request IDs.
After insertion, the switch drops the packet.
This is acceptable since a cache packet will soon service the stored request.

\textbf{Read replies.}
When receiving a read reply, the switch first checks to see if the ingress port is the recirculation port.
If it is, the reply is a cache packet.
Otherwise, it is a reply for an uncached item sent by the server.
For cache packets, if the cache misses or the state of the value is invalid, the switch drops the packet.
A cache miss for a cache packet means that the controller evicted the key from the lookup table due to a change in key popularity or the cache size.
The invalid state indicates a write request with a new value is in progress.

The switch goes through the lookup and state tables, and then looks for pending requests in the request table.
If a request is found, the switch forwards the cache packet to the client after updating the header with metadata and removing the metadata from the table slot.
To make the cache packet serve more requests, the switch clones the packet before forwarding.
The switch forwards the original packet to the client and the cloned one to the recirculation port.
The switch recirculates the packet if there are no pending requests.

\textbf{Write requests.}
The switch checks whether the requested key is in the cache lookup table.
If it is, the value for the key is invalidated to prevent reading the outdated value.
In addition, the switch sets the \texttt{FLAG} field to 1 to indicate that this request is for a cached item. 
This makes the storage server append the value in the write reply.
Regardless of a cache hit, the switch forwards the request to the storage server to update the value of the key in the server.

\textbf{Write replies.}
If the key is cached, the switch validates the value to allow read requests to get the latest value.
The switch clones the packet so that the client receives the write reply while the switch has a new cache packet simultaneously.
The switch updates the \texttt{OP} field to \texttt{R-REP} of the cloned packet after the first recirculation since cache packets should be read replies.
For a cache miss, the switch forwards the packet to the client since the reply is for an uncached item.

\textbf{Other types of messages.}
Fetch requests and correction requests are delivered to the server as standard packets.
The fetch reply is processed with the same logic as writing replies.

\begin{figure}[t!]
\centering
\includegraphics[width=8.0cm]{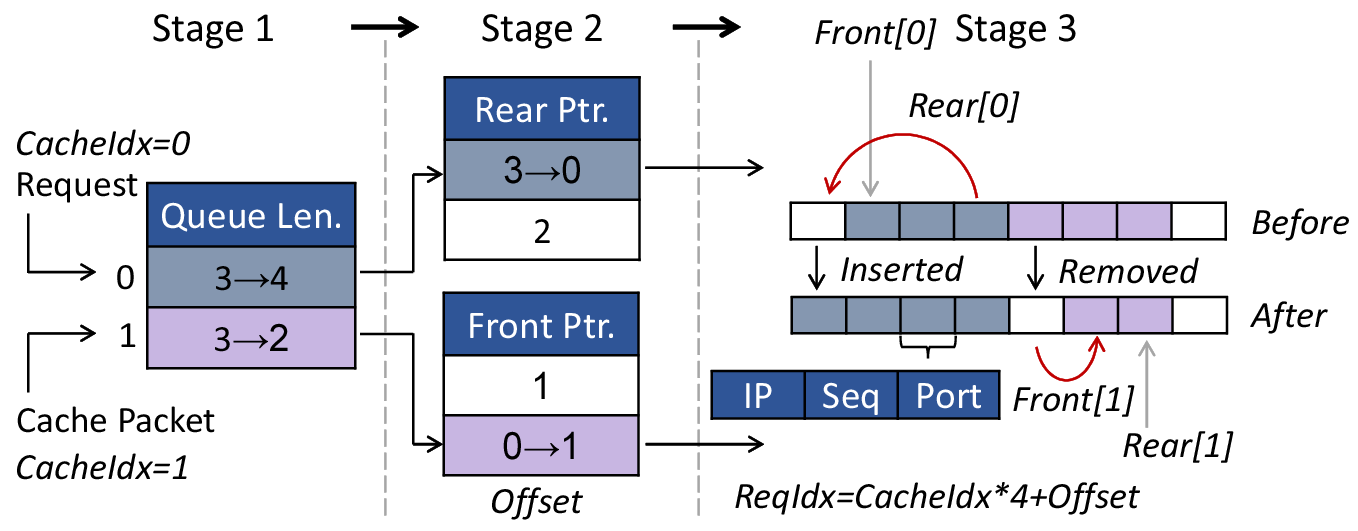}
\caption{An example of queueing operations in the request table.
\textit{
Read requests enqueue new metadata and cache packets dequeue the stored metadata.
}
\label{fig:buffering}}
\end{figure}

\subsection{Buffering Request Metadata}
\textbf{Circular queue-based request table.}
We should store multiple request metadata for cached items because the requests should wait until cache packets serve them.
The switch should be able to buffer concurrent read requests for the same cached item.
If not, many requests for the cached item would be forwarded to the storage server due to the lack of vacancies.
Furthermore, a request for a key should be isolated from requests for other keys since the request may undergo excessive queueing delay due to the different requests.

We design a request table based on a circular queue structure.
Our request table provides a logical queue for each cached key, and the queue can be accessed in $O(1)$ by efficient indexing.
This provides fast and isolated queue access for a key regardless of other keys.
The request table consists of 6 register arrays: the first three arrays are for storing request metadata for each key (i.e., client IP address, request sequence number array, and L4 port number).
The other three arrays are for managing queue operations.
The queue length array maintains the number of stored request metadata for each key.
The front pointer array handles dequeue operations by tracking the first request metadata for each key.
The rear pointer arrays perform enqueue operations by keeping track of the last request metadata for each key.

The queue management arrays are indexed using a table index ($CacheIdx$) returned by the cache lookup table.
The request metadata arrays are accessed via a request index, which is defined as $ReqIdx = CacheIdx\times S + i$ where $S$ is the maximum queue size per key and $i$ is the offset for $i$th slot in the logical queue for a key and is given by the pointer arrays.
The switch uses three match-action stages for a request table.
Specifically, the switch first checks the queue status (Stage 1), and performs en/dequeue operations if the queue is not full/empty (Stage 2).
Next, the switch gets/puts metadata from/into the request table (Stage 3).

\textbf{An example for table operations.}
Figure~\ref{fig:buffering} shows a simple example of table operations.
We assume that the request table can maintain up to 4 requests for each key.
We consider enqueueing and dequeueing operations using read requests and cache packets, respectively.
In stage 2, the rear pointer changes to 0 from 3 since we implement a circular queue.
It is also easy to see that the request metadata for different keys does not collide since we partition the metadata arrays using the indexing formula of $ReqIdx=CacheIdx\times S + i$.

\subsection{Cache Serving}
\textbf{Handling multiple requests via packet cloning.}
In \sys, cache packets pass through the switch data plane repeatably to serve pending requests for cached items.
One challenge is serving multiple requests with a cache packet fetched only once because the cache packet is forwarded to the client after updating the header using the retrieved metadata.
A strawman is to fetch the cache packet from the server again, but this approach is inefficient as the switch cannot serve pending requests for the key until the fetching is completed.
Fetching multiple copies of the cache packet may be a solution, but this incurs excessive queueing delay between cache packets. 
In addition, it is hard to know how many cache packets should be fetched because we cannot predict the number of requests for each cached item in advance.

\begin{figure}[t!]
\centering
\includegraphics[width=8.0cm]{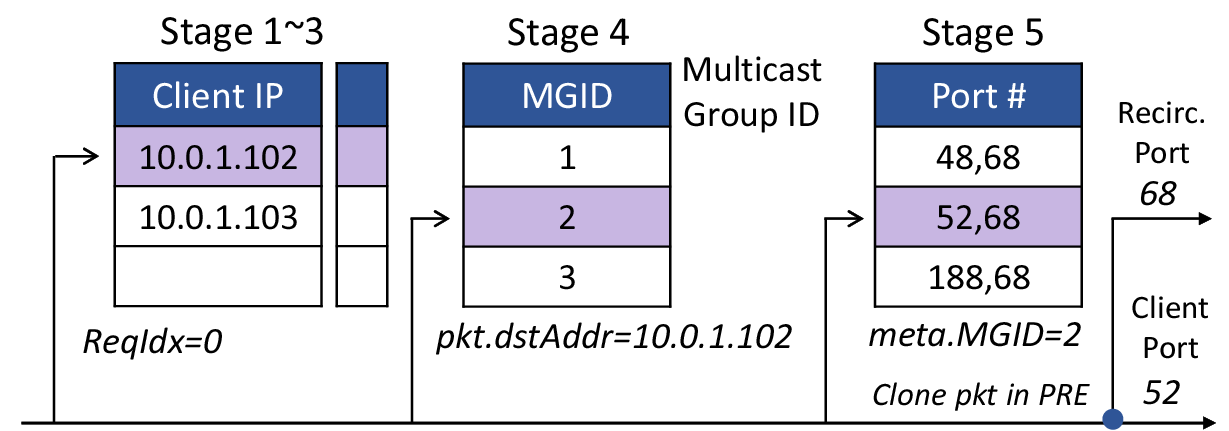}
\caption{An example of cache packet cloning using multicast.
\textit{
Packet cloning is done in the PRE, a hardware module of the switch ASIC for packet cloning.
}
\label{fig:serving}}
\end{figure}

To address this, we utilize packet cloning, another built-in feature of the programmable switch ASIC besides packet recirculation.
Packet cloning is done by the PRE, a hardware module specialized for packet cloning in the switch ASIC.
Cloning has low overhead for the following reasons.
First, the PRE is located after the ingress pipeline.
This means the switch does not repeat the ingress pipeline processing for the cloned packet, not causing extra ingress processing delay.
Second, the switch does not copy the entire packet.
It only copies the small descriptor pointing to the memory location of the packet and reuses the packet data.

We use multicast to forward the original and clone packets.
Specifically, through a match-action table using the destination IP address acquired from the request table, the switch gets a multicast group ID that specifies the regular port number directed to the client and the internal recirculation port number.
The switch forwards the original packet to the client and the cloned one to the switch data plane again, making it serve more requests.

\textbf{An example of cache cloning.}
In Figure~\ref{fig:serving}, a cache packet gets the client IP address of 10.0.1.102 and other metadata from the request table.
The packet obtains the multicast group ID as 2 using the destination IP address (i.e., the client IP address).
The cache packet then gets a pair of destination port numbers.
52 is the port number directed to the client, and 68 is the recirculation port number.
After that, the switch clones the packet and forwards the original and cloned ones to each port.

\begin{figure}[t!]
\centering
\includegraphics[width=8.0cm]{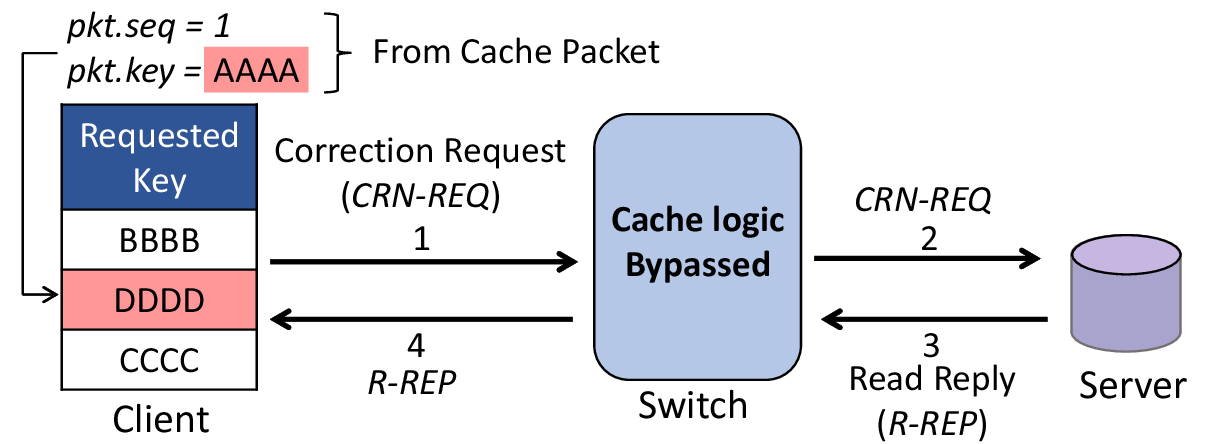}
\caption{An example of handling hash collisions.
\textit{
The client gets the correct value by sending a correction request.
}
\label{fig:hashcollision}}
\end{figure}

\subsection{Handling Hash Collisions}
The existing works use the item key as the match key of the cache lookup table, which is implemented as a match-action table.
Since the match-key width is limited by hardware, we cannot use the key exceeding the size limit.
We may use a register array, but the size limitation in the register index is stricter than in the match-action table.
Therefore, we use the fixed-sized key hash as the match key.
A challenge is how to resolve potential hash collisions.
We should handle this because a request may read the wrong value of other keys.

\textbf{Client-side collision resolution.}
We handle hash collisions at the client level by maintaining a list of the keys for each request that has not yet received a reply.
The list is indexed by $pkt.seq$, a request ID.
Request packets contain both the original key and the key hash since we should use the original key to get the value in servers.
Upon receiving read replies, the client checks whether the requested key in the list and the returned key in the reply header are identical.
If different, the client sends a correction request to the storage server.
The switch then forwards the request without applying the cache logic so that the client gets the correct item value from the storage server.
One issue is that requests for uncached but hash-collided keys always undergo this process.
However, this is uncommon since our 128-bit key hash provides a low collision probability. 
In our experience, hash collisions occasionally occur during cache updates.

\textbf{An example of resolving hash collisions.}
We plot an example of handling hash collisions in Figure~\ref{fig:hashcollision} where the requested key is \texttt{DDDD} but the returned key is \texttt{AAAA}.
Upon receiving the reply, the client detects that the retrieved value is a wrong value by referring to the key list using $pkt.seq=1$.
The client then sends a correction request to the storage server.
The switch bypasses the cache logic, and forwards the packet to the server.
The storage server returns the value as a read reply, and the client finally gets the correct value for the key \texttt{DDDD}.
The client removes the key from the list.
$pkt.seq$ wraps around if it reaches the maximum value.

\subsection{Cache Coherence}
To ensure cache coherence between the switch and servers, we design an invalidation-based coherence protocol illustrated in Figure~\ref{fig:pro} (c) and (d).
The switch invalidates the value when handling a write request for a cached item and revalidates the value upon receiving a write reply.
Ideally, the outdated cache packets should be replaced with the latest cache packet.
However, requests may be returned with the stale value since the outdated packets are still circulating in the switch data plane even if the item value becomes invalid.
To address this, the switch drops the cache packet if the item is cached but its value is invalid.
Owing to the fact that the cache packet is dropped before accessing the request table, we can prevent read requests from retrieving the stale value until a new cache packet is fetched.
The storage server sends a single reply packet, and the switch updates the value and replies to the client simultaneously by cloning the packet.

\begin{figure}[t!]
\centering
\includegraphics[width=8.0cm]{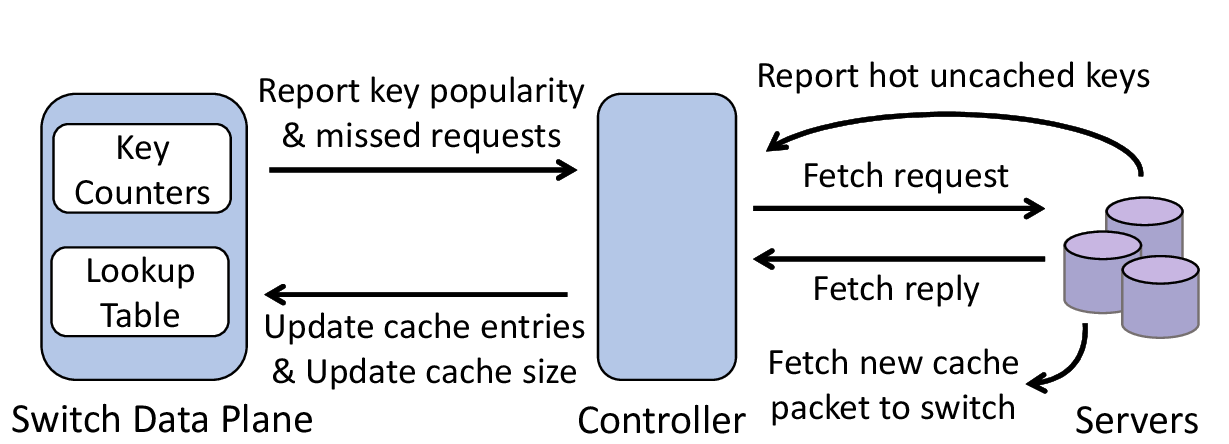}
\caption{Cache update. 
\textit{
The controller in the switch control plane updates the entries of the cache lookup table and the cache size by tracking workload changes.
}
\label{fig:update}}
\end{figure}

\subsection{Cache Updates}
\sys can handle dynamic workloads by tracking workload changes as follows.
To handle key popularity changes over time, we update cache entries based on periodic popularity reports from the switch and storage servers.
In particular, we count the popularity of cached keys in the switch data plane using the key popularity counter.
The controller keeps track of key popularity by reading the counter periodically.
Meanwhile, storage servers periodically report the top-$k$ keys to the controller, which are popular uncached keys.
The servers use a count-min sketch with five hash functions to track key popularity in a memory-efficient manner while ensuring accuracy.
To reflect the recent status only, we reset all the counters to zero after reporting.

If the cache needs to be updated due to a change in key popularity, the controller deletes the victim key from the cache lookup table and inserts a new popular key instead.
The controller then sends a fetch request to the storage server containing the latest value for the key.
Upon receiving the request, the server fetches a new cache packet to the switch data plane and replies to the controller.
The new popular key inherits the table index ($CacheIdx$) of the evicted key.
With this, the pending requests for the evicted key can be handled by the new cache packet and the hash collision resolution mechanism.
Note that the hash collision here is due to the same table index, not that the hashes actually collide.

\subsection{Handling Practical Requirements}
\textbf{Failure handling.}
For packet loss, we can use an application-level mechanism.
Our controller uses UDP with a timeout-based mechanism to exchange fetch requests/replies.
Storage servers use TCP to report top-$k$ items to the controller.
Meanwhile, server failures do not cause a \sys-specific problem.
Switch failures result in the loss of cached items, but the cache can be reconstructed quickly by the controller after the switch is recovered because the switch failure is similar to the rapid key popularity changes.

\textbf{Multi-rack deployment.}
\sys can be deployed for clusters where multiple racks exist because the ToR switch caches hot items of storage servers belonging to its rack only.
For example, assume that we have two racks where ToR switches, $ToR1$ and $ToR2$, are interconnected via a spine switch $SPN$.
When $CLI$, a client in rack 1, sends a request to $SRV$, a server in rack 2, the path for an uncached item is $CLI-ToR1-SPN-ToR2-SRV-ToR2-SPN-ToR1-CLI$.
If the requested item is cached, $ToR2$ is the only switch that applies the cache logic.
There is no issue even if rack 1 is the replicated rack of rack 2 where the two ToR switches have the same cache entries because the request is handled by $ToR1$ directly.
If the client is located in another rack and the other two racks are replicated ones, we can balance requests for those racks by implementing load-balancing mechanisms in the ToR switch of the rack where the client exists.

\subsection{Discussion}
\textbf{Dynamic cache sizing.}
The value size of hot items may change over time, or the value size of a new hot item may differ from that of the evicted item.
In this case, the optimal cache size can be changed.
To address this, the controller may change the cache size to maximize the performance.
The controller calculates the request overflow ratio periodically, which is defined as the number of requests for cached items that were sent to the server divided by the total number of requests for cached items.
The controller compares the ratio with a pre-defined threshold (e.g., 1\%).
If the ratio exceeds the threshold, the controller decreases the cache size.
Otherwise, the cache size is increased.
We can also set the minimum and maximum cache sizes to prevent overreaction.

\textbf{Multi-packet items.}
Many key-value items are less than the MTU size.
Therefore, existing works in key-value stores generally consider item values up to 1024 bytes~\cite{switchkv,mica,liu17, leed,fawn,kvell}.
However, some items may exceed the MTU size, such as articles and photo objects for some workloads (e.g., Wikipedia and Flickr~\cite{blott15}).
To cache multi-packet items, we should fetch multiple cache packets with fragmented values for the same key.
To do this, we maintain the number of forwarded cache packets for each item by placing another register array alongside the request table, the \textit{ACKed packet counter}.
The initial value of each slot is 1 since most items are single-packet.
When fetching item values, the storage server puts the number of packets comprising the item in the \texttt{FLAG} field.
When handling cache packets, the switch retrieves metadata as usual, but it does not manipulate the slots of all register arrays in the request table if the slot value in the ACKed packet counter is not equal to the value in \texttt{FLAG}.
Instead, the switch increases the slot value by one.
When the slot value in the counter equals \texttt{FLAG}, the switch removes request metadata as it is ready to be finished.

However, this may reduce the performance gains since multiple cache packets occupy the cache for one large item.
Therefore, the balancing efficiency may decrease if many items consist of multiple packets.
One alternative direction is designing a hierarchical cache architecture.
We may integrate \sys with Pegasus~\cite{pegasus} where the switch replicates hot items across storage servers.
Specifically, we can use the switch as the primary cache and the storage servers (or additional cache servers) as secondary caches.
For example, we can cache hot single-packet items in the switch while caching hot multi-packet items to servers.

\textbf{Write-back caching.}
Although most workloads are read-intensive, some workloads are write-intensive.
Since we use write-through caching that updates the switch and the server simultaneously, our performance gain decreases as the write ratio grows.
FarReach~\cite{farreach} is a recent work that enables write-back caching.
Although it still cannot serve many workloads due to the size limitation, it provides high performance regardless of the write ratio.
The difference between write-through caching and write-back caching is whether a write request for cached items updates the storage server or not.
Therefore, \sys can use write-back caching as well by letting the switch return write replies upon receiving write requests after updating the cache only, though we need extra modules like snapshot generation.

\textbf{Multi-pipeline deployment.}
The multi-terabits performance of the programmable switch ASIC comes from the multi-pipelined architecture.
Each pipeline consists of tens of regular ports and one internal recirculation port.
The switch resources are partitioned, and metadata and memory data are not shared between the pipelines.
Therefore, the pipelines of client-directed ports and server-directed ports should be the same.
Otherwise, cache packets in a pipeline may not read requests since they are in another pipeline.
This can be addressed by mapping each pipeline of the ToR switch to a client-directed port on the spine switches.
This is especially feasible when the clients and servers are located in different racks.
We expect this to be addressed more easily with a new programmable switch architecture like MP5~\cite{mp5} that achieves high performance with a logical single pipeline.
\section{Implementation\label{implementation}}
\textbf{Client-server application.}
We develop an open-loop application in C using NVIDIA Messaging Accelerator library (VMA)~\cite{vma}.
VMA bypasses kernel network stacks by intercepting the socket function calls and translating them to native RDMA verbs.
The client application measures throughput and latency by generating requests.
The time gap between consecutive requests follows an exponential distribution.
The server application has multiple threads where each thread is pinned to a disjoint CPU core.
To emulate multiple storage servers, we assign a partition per thread so that each thread acts as an independent storage server.
We also limit the Rx throughput of each emulated server to 100K RPS to ensure the bottleneck is at servers in our testbed.
This technique is used in existing works~\cite{netcache,switchkv,farreach} as well.
In a similar vein, like \sota~\cite{netcache}, we implement a key-value store with TommyDS~\cite{tommyds}, a high-performance hash table library.

\textbf{Switch.}
We implement the switch data plane in P4$_{16}$~\cite{bosshart14} for Intel Tofino~\cite{tofino}.
We use Intel P4 Studio SDE 9.7.0 to compile the switch data plane.
Our prototype uses 9 stages and 6.67\% SRAM, 7.38\% Match Input Crossbar, 9.29\% Hash Bit, and 30.56\% ALUs.
The request table has a maximum queue size of 8 for each key.
The controller is written in Python 3.
In our prototype, we use an additional register array for the request table to store the timestamp of the request for latency measurement.
The \sys header has 3 extra fields of 1-B \texttt{Cached}, 4-B \texttt{Latency}, and 1-B \texttt{SrvID}.
The first two fields are required to separately measure the latency of requests for cached and uncached items.
The last field is to store the emulated server ID as we emulate multiple storage servers as dedicated threads in a physical node.
\section{Evaluation\label{evaluation}}
\subsection{Methodology}
\textbf{Testbed setup.}
We build a cluster consisting of 8 nodes, which are connected by an APS BF6064X-T switch with the Intel Tofino 1~\cite{tofino}.
The servers are equipped with a 10-core CPU (Intel i5-12600K @ 3.7 Ghz, 12 hyperthreads and 4 non-hyperthreads), 32 GB of DDR5 memory, and a 100GbE NVIDIA CX-5 NIC.
The servers run Ubuntu 22.04 LTS with Linux kernel 6.5.0.
The 4 nodes act as clients, and the remaining 4 nodes are used to emulate multiple storage servers.

\textbf{Compared schemes.}
We compare our work with \baseline and \sota~\cite{netcache}.
\baseline is a mechanism without any cache logic.
\sota is the representative in-network caching architecture.
FarReach~\cite{farreach} and DistCache~\cite{distcache} also adopt the \sota architecture.
We implement the core caching logic of \sota, but our implementation provides items up to 64-byte values across 8 stages with an 8-byte accessible size per stage.
We find that the P4 compiler allocates only two cache read tables per stage, even with pragma statements.
We suspect that this is due to compiler restrictions on our code.
We clarify that this does not mislead the conclusions of our experiments, as there is a negligible difference in the latency required to read 64 bytes and 128 bytes at line rate using the same number of match-action stages.

\textbf{Workloads.}
We emulate a single storage rack with 32 storage servers using 8 partitioned threads per node.
We basically consider a workload with 10M key-value pairs whose key popularity skewness follows a Zipfian distribution with $\alpha=0.99$, since it is regarded as typical skewness~\cite{ycsb,atikoglu12}.
Although both key and value sizes impact whether an item is cacheable by \sota, we use 16-byte keys by default for simplicity, the maximum supported key size by \sota.
Instead, we represent the portion of cacheable items using different ratios between 64 bytes and 1024 bytes values.
The 64-byte value represents a cacheable item value of \sota.
The 1024-byte value is a typical-sized item value in many workloads and is considered in many research papers~\cite{switchkv,mica,liu17, leed,fawn,kvell}.
We use a bimodal distribution with 82\% 64-byte and 18\% 1024-byte values by considering the cacheable item ratio of \sota for the \texttt{Cluster018} workload of Twitter~\cite{cao20}.
Most experiments are for read-only workloads as we target read-intensive workloads.

Except for dynamic workloads, we preload the 10K and 128 hottest items for \sota and \sys, respectively.
128 is a nearly optimal cache size for \sys that provides the best performance gains.
Since \sota can cache only 82\% of items in the workload, the actual number of cached items is 8200, which is still larger than the cache size of \sys by 64$\times$.
To be fair, we choose 82\% of keys among the 10K hottest keys with a uniform distribution. 
We store the chosen keys as a text file to make experimental results consistent.
The controller loads the file and puts the keys to the switch cache.
We note that, even if we use different key samples, trends have been consistent.

\subsection{Main Results}
\begin{figure}[t!]
\centering
\includegraphics[width=7.0cm]{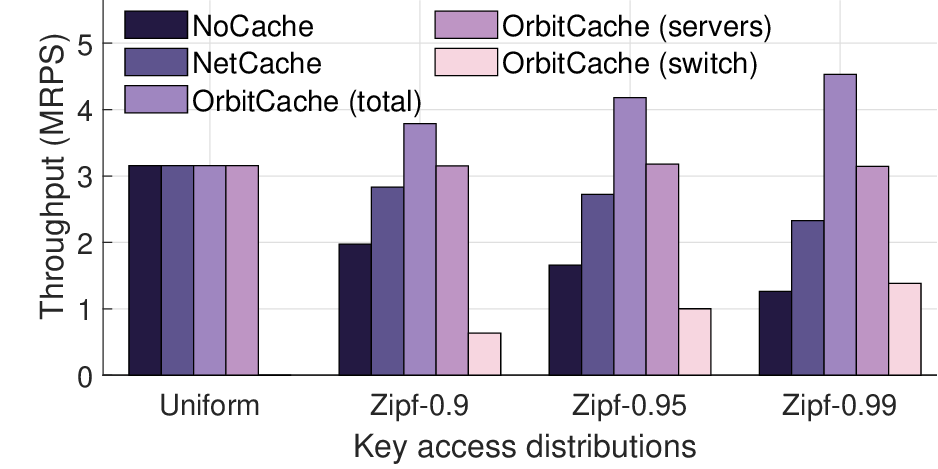}
\caption{Throughput with different skewness.
\label{fig:thdist_main}}
\end{figure}

\textbf{Throughput with different key access distributions.}
We measure the throughput with different skewness and plot the results in Figure~\ref{fig:thdist_main}.
We can see that \sys provides high throughput regardless of skewness, unlike the others whose throughput decreases as the workload becomes more skewed.
\sota does not provide high throughput as expected since many hot items are not cacheable.
In the Zipf-0.99 workload, \sys has higher throughput than \baseline and \sota by 3.59$\times$ and 1.95$\times$, respectively.
The server throughput in \sys is consistent across the given skewness, and this means that the loads are balanced.

\begin{figure}[t!]
\centering
\includegraphics[width=7.0cm]{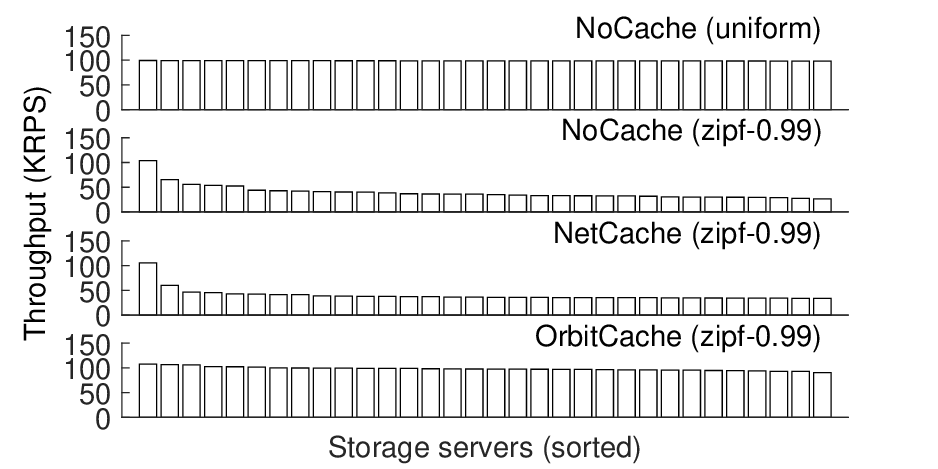}
\caption{
Load on individual storage servers.
\label{fig:imbalance_main}}
\end{figure}

\textbf{Individual server loads}.
We plot the load on individual servers in Figure~\ref{fig:imbalance_main}.
We can see that \baseline and \sota do not balance loads well.
This is due to the lack of caching logic for \baseline and many uncacheable hot items for \sota.
However, \sys can balance the loads since \sys can cache variable-length items.

\begin{figure}[t!]
\centering\hfill
\subfloat[Median]{\includegraphics[width=0.480\linewidth]{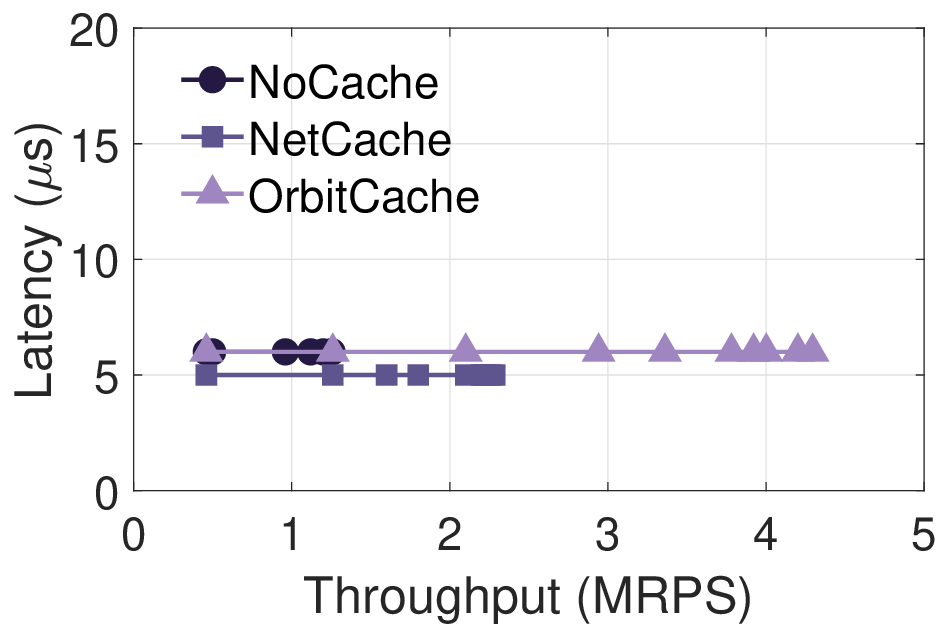}}\hfill
\subfloat[99th percentile]{\includegraphics[width=0.480\linewidth]{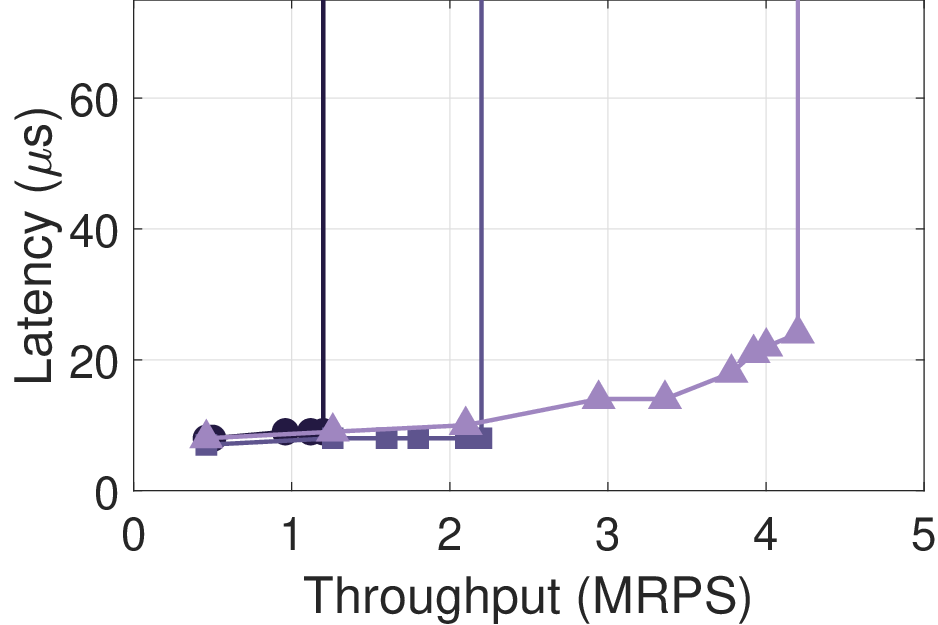}}\hfill
\caption{Latency vs. throughput.
\label{fig:throughputlatency}}
\end{figure}

\textbf{Latency vs. throughput.}
We measure the latency by varying Tx throughput.
Figure~\ref{fig:throughputlatency} shows the median and the 99th percentile latencies as a function of Rx throughput.
\sys provides the best throughput but slightly higher latency than \sota of 1 microsecond.
This is because \sota has a larger cache size, and requests handled by the switch dominate in latency data.
In addition, in \sys, requests for cached items should wait until cache packets read them. 
Although \sota has slightly better latency, throughput is very limited since it fails to balance loads.

\begin{figure}[t!]
\centering
\includegraphics[width=7.0cm]{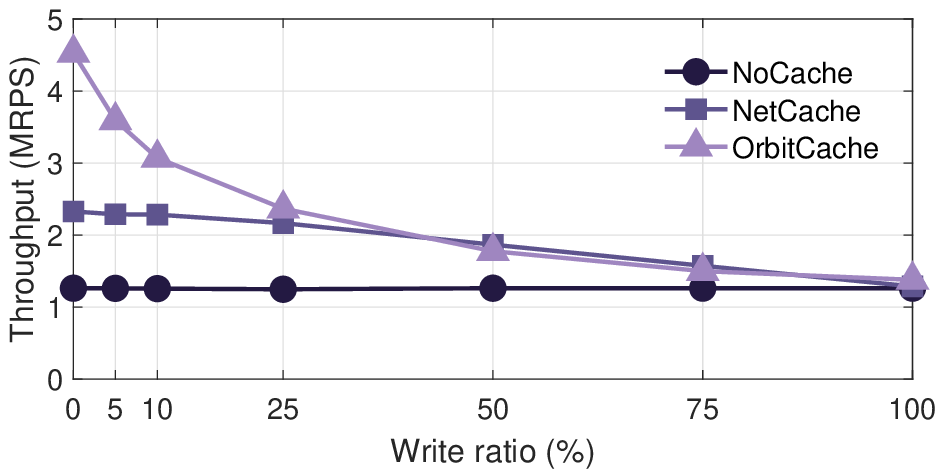}
\caption{
Impact of write ratio.
\label{fig:impactofwrites}}
\end{figure}

\textbf{Impact of write ratio.}
Figure~\ref{fig:impactofwrites} reports the throughput as a function of write ratios.
The throughput of \sys decreases as the write ratio grows.
This is because the switch invalidates the cached key when handling a write request for a cached item to ensure cache coherence.
The switch cache provides performance gains for read requests, the throughput is decreased with higher write ratios since read requests for invalid cached items are forwarded to storage servers.
With the 100\% write ratio, the throughput of \sys converges to \baseline since the cache does not provide any benefit.
Meanwhile, similar to \sys, \sota offers reduced throughput as the write ratio grows since it also invalidates the cached key if there is a write for the key.

\begin{figure}[t!]
\centering\hfill
\subfloat[Throughput]{\includegraphics[width=0.480\linewidth]{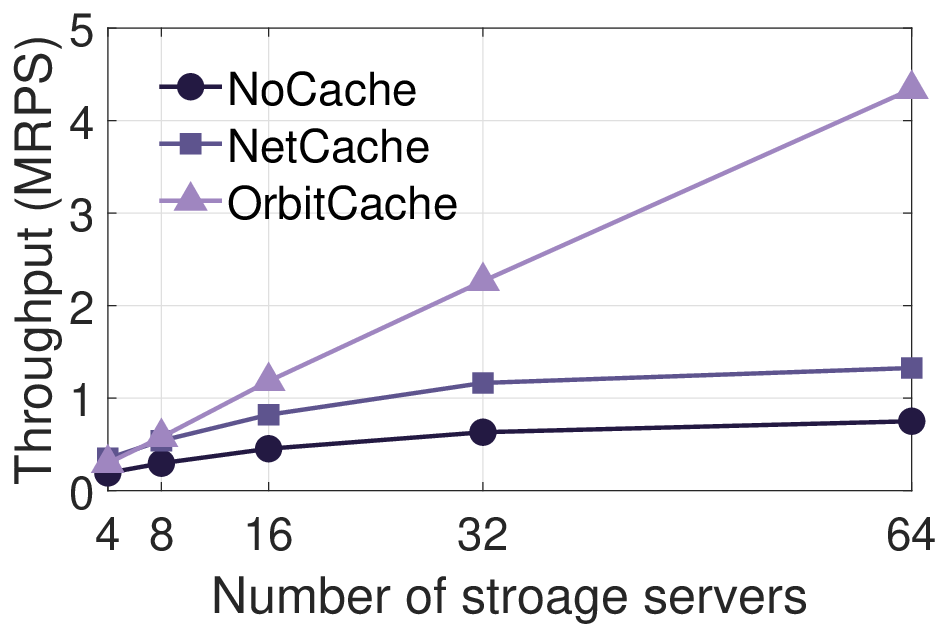}}\hfill
\subfloat[Balancing efficiency]{\includegraphics[width=0.480\linewidth]{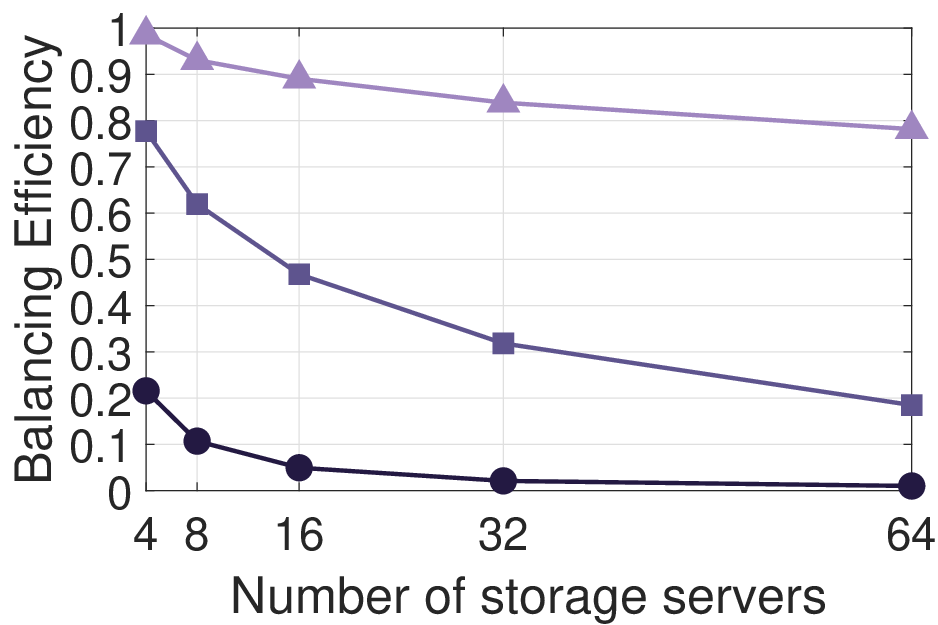}}\hfill
\caption{Scalability.
\label{fig:scalability}}
\end{figure}

\textbf{Scalability.}
We now inspect whether \sys can balance loads of different numbers of servers.
In this experiment, we limit the Rx throughput to 50K RPS to ensure that the bottleneck occurs at the storage servers rather than the clients, even when using 64 servers.
In Figure~\ref{fig:scalability} (a), we can see that the throughput of \sys is improved almost linearly while the others do not.
Figure~\ref{fig:scalability} (b) clarifies that this is because \baseline and \sota fail to balance imbalanced loads.
\textit{Balancing efficiency} is defined as the minimum throughput between the servers divided by the maximum throughput between the servers.

We note that \sys may not balance workloads with larger servers like hundreds of servers because the cache size is limited.
\sys targets a single storage rack consisting of tens of servers, and our experiment result demonstrates that \sys is enough to balance a rack-scale storage system.
To scale out for hundreds of servers, instead of relying on a single switch to balance all servers, each ToR switch in clusters should cache items of storage servers belonging to its rack, as described in Section 3.9.

\textbf{Performance with production workloads.}
We conduct experiments with several workloads of Twitter~\cite{yang20} to see how \sys works with various production workloads.
We pick 5 workloads based on the cacheable item ratio.
We assign the workload ID A to E for \texttt{Cluster045/016/044/017/020}.
We still use the 16-B keys for simplicity but vary the write ratio, the portion of 64-B values.
Unlike the other experiments, the cacheable item ratio is controlled by choosing keys with a uniform distribution independent of the portion of small 64-B values.

\begin{figure}[t!]
\centering
\includegraphics[width=7.0cm]{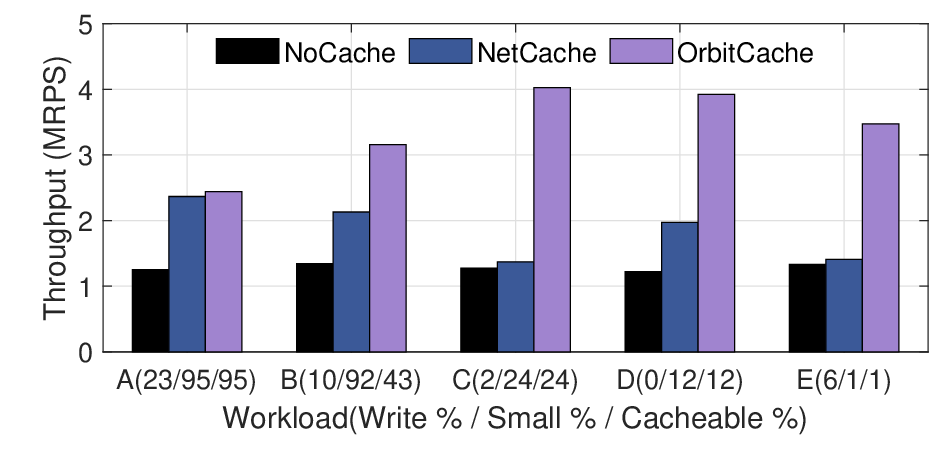}
\caption{
Performance with production workloads.
\label{fig:impactofworkload}}
\end{figure}

Figure~\ref{fig:impactofworkload} shows the results.
Although the difference in throughput varies depending on the used workloads, we can see that \sys shows the best performance for all the workloads.
There is a little difference for Workload A.
This is because \sota can cache 95\% of items, and the write ratio is relatively high.
Workload E is a workload that shows a significant performance gap.
This is because only 1\% of items are cacheable for \sota.

\subsection{Deep Dive}

\begin{figure}[t!]
\centering\hfill
\subfloat[Median]{\includegraphics[width=0.480\linewidth]{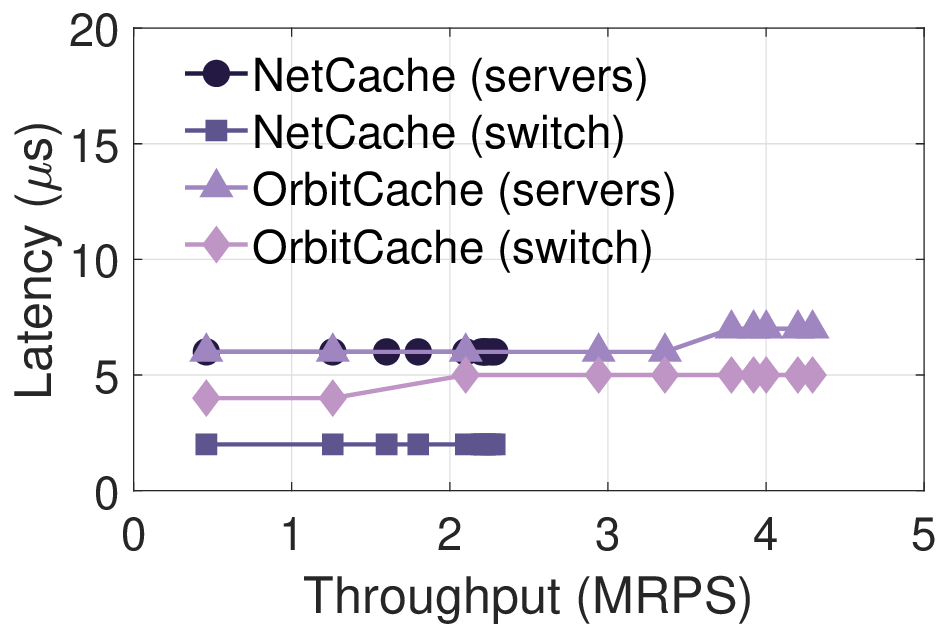}}\hfill
\subfloat[99th percentile]{\includegraphics[width=0.480\linewidth]{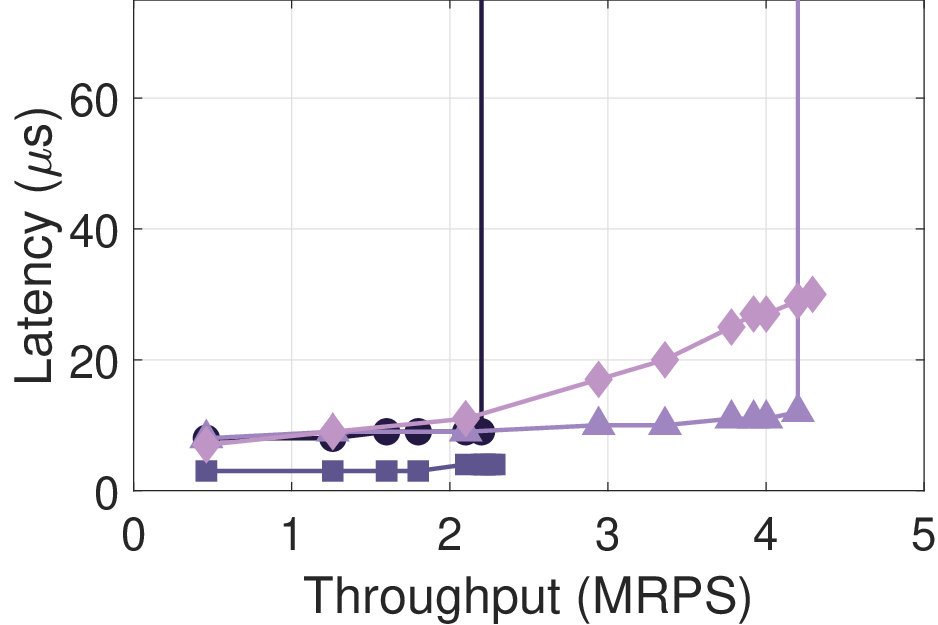}}\hfill
\caption{Latency breakdown.
\label{fig:latency_breakdown}}
\end{figure}

\textbf{Latency breakdown.}
Figure~\ref{fig:latency_breakdown} (a) and (b) plot the median and the 99th percentile latency breakdowns, respectively.
In the median latency, \sys shows slightly higher switch latency than \sota.
This is because requests in \sys should wait until circulating cache packets read them, resulting in latency overhead.
In Figure~\ref{fig:latency_breakdown} (b), we can see that the tail latency of \sys (switch) increases as throughput grows.
This is not surprising because we use a circular queue-based request table to maintain request metadata and packet cloning for a cache hit, resulting in additional latency overhead.
We believe that this is acceptable since \sys can provide much higher throughput than the others.
In addition, the tail latency of the switch is still tens of microseconds even when the tail latency of servers soars by reaching the saturated throughput.
We can decrease the tail latency of the switch by reducing the cache size as well.
If the cache is based on a commodity server, the tail latency would be 10-100$\times$ longer than the switch cache.

\begin{figure}[t!]
\centering\hfill
\subfloat[Saturated Throughput]{\includegraphics[width=0.480\linewidth]{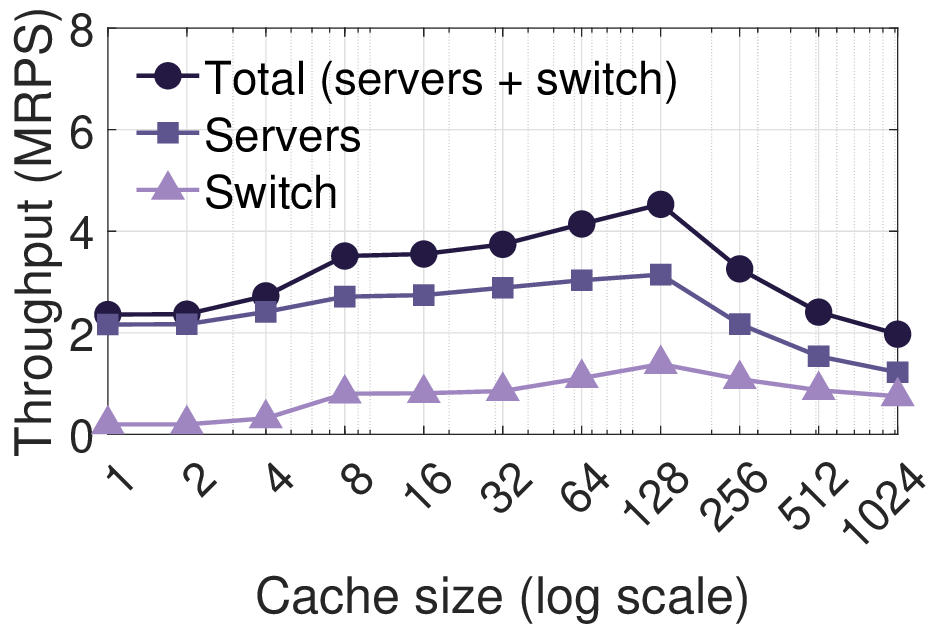}}\hfill
\subfloat[Switch Cache Latency]{\includegraphics[width=0.480\linewidth]{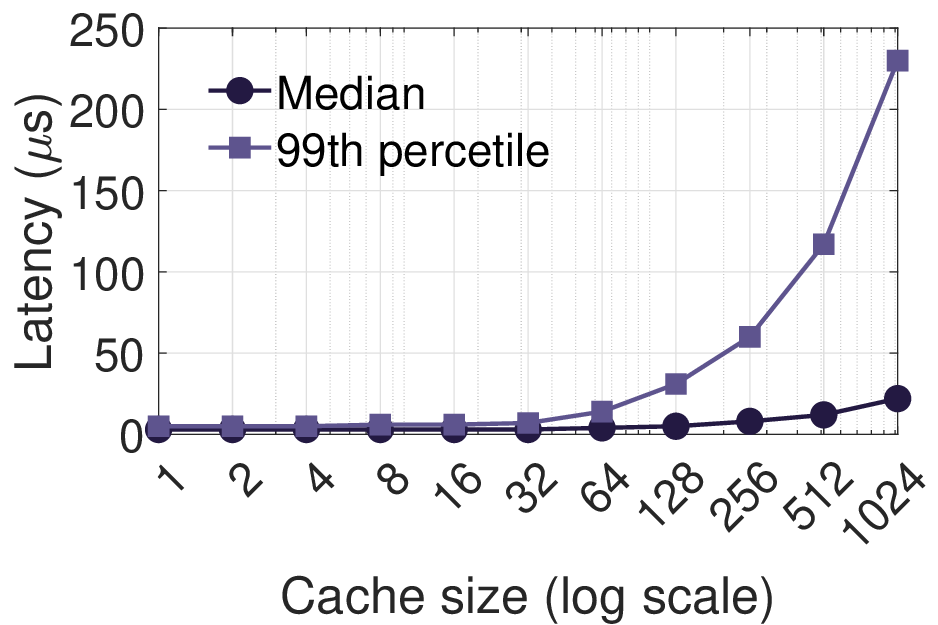}}\hfill
\subfloat[Overflow request ratio]{\includegraphics[width=0.480\linewidth]{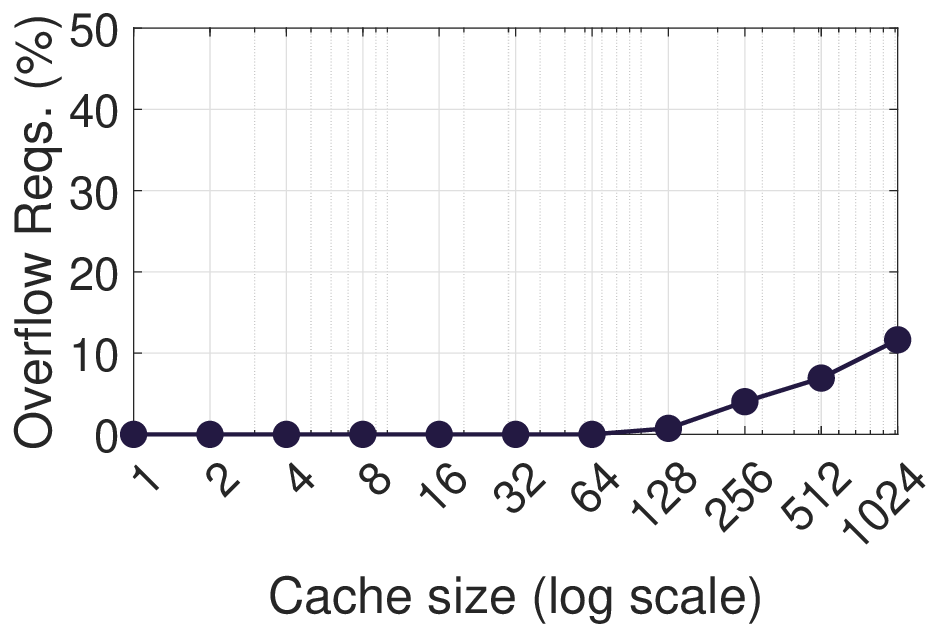}}\hfill
\caption{
Impact of cache size.
\label{fig:impactofcachesize}}
\end{figure}

\textbf{Impact of cache size.}
Figure~\ref{fig:impactofcachesize} plots the throughput breakdown, request latency handled by the switch, and the overflow request ratio of \sys with different cache sizes.
We can see that the total throughput increases as the number of cached entries grows.
However, the throughput is saturated with around 128 cached items.
Similarly, in Figure~\ref{fig:impactofcachesize} (b), the tail latency increases quickly after 64-128 cached items.
Figure~\ref{fig:impactofcachesize} (c) clarifies the reason.
From 256 cached items, the overflow request ratio rapidly increases, where the overflow requests indicate the requests for cached items forwarded to the storage servers due to the lack of free slots in the request table.
The lack of free slots is caused by the excessive queueing delay between too many cache packets beyond the capability of the switch hardware.
This clarifies that the cache size is the key to determining the performance.
This also means we should choose an effective cache size between 32 and 128 to balance the trade-off.

\begin{figure}[t!]
\centering
\subfloat[Throughput]{\includegraphics[width=0.70\linewidth]{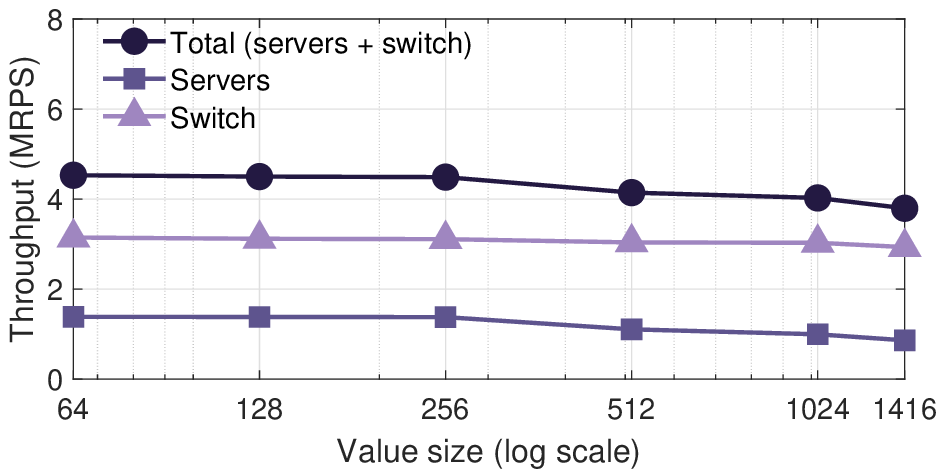}}\\
\subfloat[Balancing efficiency]{\includegraphics[width=0.480\linewidth]{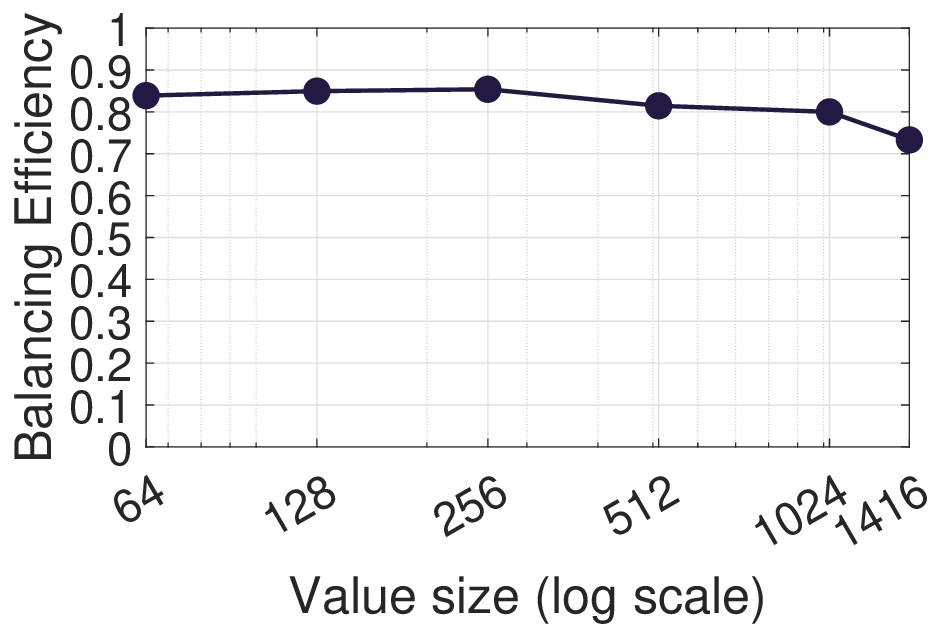}}\hfill
\subfloat[Cache size]{\includegraphics[width=0.480\linewidth]{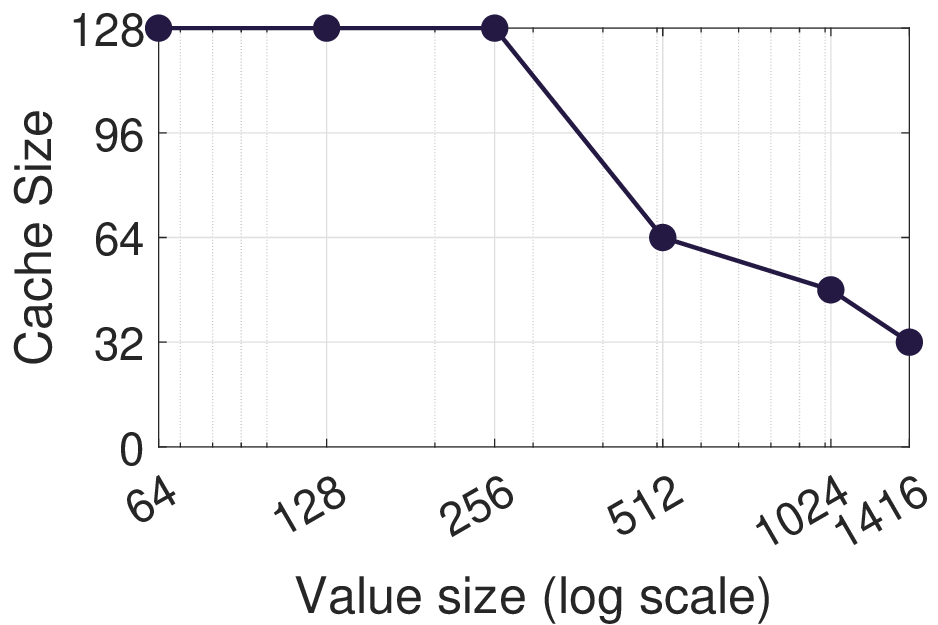}}\hfill
\caption{Impact of item size.
\label{fig:impactofvaluesize}}
\end{figure}

\textbf{Impact of item size.}
We measure throughput, balancing efficiency, and the maximum cache size by varying the item size.
In this experiment, we use the 100\% same value size for all items to inspect how \sys works in the worst case.
Note that 16-B key and 1416-B value are the maximum size for a single packet payload with 28-B custom header fields.

Figure~\ref{fig:impactofvaluesize} (a) shows that \sys can balance various workloads, including a workload where 100\% items are MTU-sized.
The slight drops in throughput are due to the increased value size.
Figure~\ref{fig:impactofvaluesize} (b) shows the balancing efficiency.
We see that \sys generally maintains high balancing efficiency.
Figure~\ref{fig:impactofvaluesize} (c) plots the effective cache size that provides the highest performance gain.
We can see that the effective cache size decreases as the value size increases.
This is because larger cache packets consume more switch resources, making the effective cache size small.
Nevertheless, we can see that \sys can balance the workloads well since we need only a small number of cached items to balance skewed workloads.
Due to the page limit, we omit the result in the impact of key size, which is similar to the result in the impact of value size.

\textbf{Handling dynamic workloads.}
We investigate how \sys reacts to dynamic workloads.
Like existing works~\cite{netcache,farreach,switchkv}, we use a \textit{hot-in} pattern, which is the most radical workload change.
We use 4 storage servers without server emulation and Rx rate limits similar to a previous work~\cite{farreach} to avoid inaccuracy due to the system state change.
Here, every 10 seconds, the popularity of the 128 coldest items and the 128 hottest items is swapped.
This means that all cache entries should be replaced suddenly.
These sudden popularity changes may not be common in practice, but this enables us to evaluate the robustness of \sys.

\begin{figure}[t!]
\centering\hfill
\subfloat[Throughput]{\includegraphics[width=0.480\linewidth]{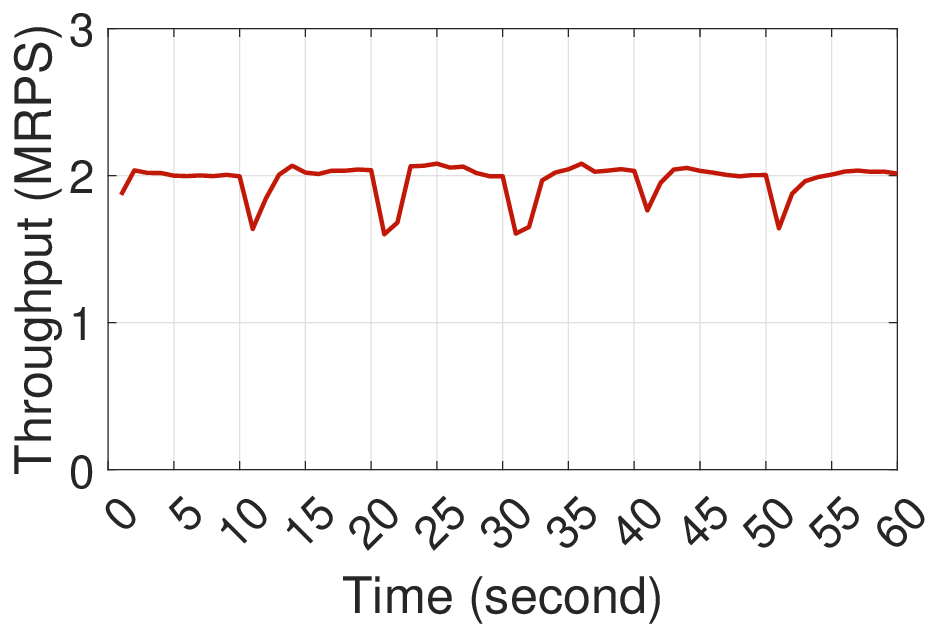}}\hfill
\subfloat[Overflow request ratio]{\includegraphics[width=0.480\linewidth]{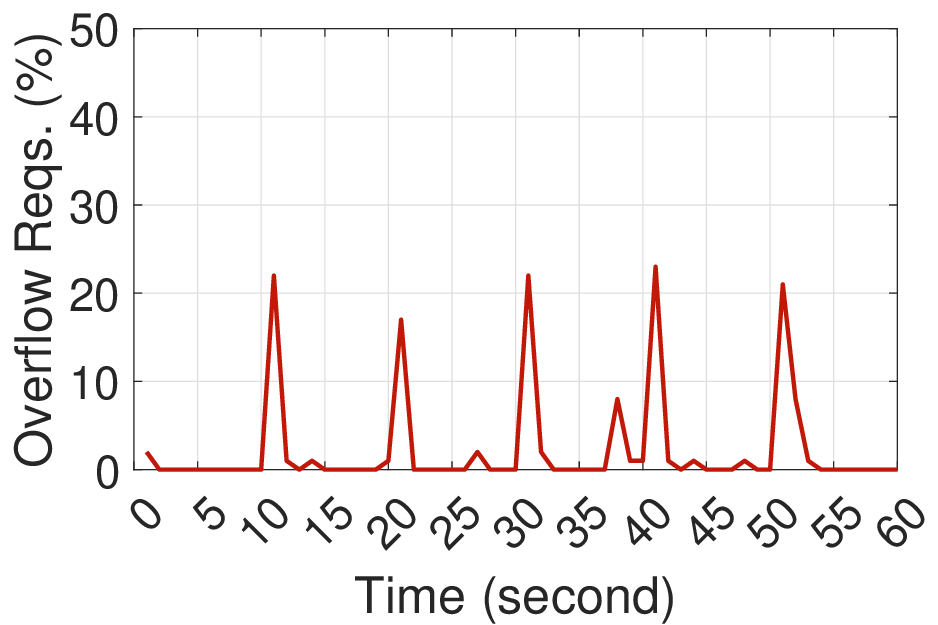}}\hfill
\caption{
Performance with dynamic workloads.
\label{fig:dynamic}}
\end{figure}

Figure~\ref{fig:dynamic} (a) shows the throughput for 60 seconds.
We can see that the throughput decreases when key popularity changes but recovers within a few seconds.
This is because the controller in the switch control plane quickly updates the cache entries and fetches cache packets based on the top-$k$ report of servers and the cache popularity counter of the switch data plane.
Figure~\ref{fig:dynamic} (b) plots the change in the overflow request ratio over time.
We can see that the ratio soars when the key popularity changes.
This is because it takes time to fetch items from storage servers.
Similar to the throughput result, the overflow ratio decreases after the items are fetched.
It would be interesting if we could design a mechanism that minimizes the cache update time.

\section{Related work\label{relatedwork}}
\textbf{In-network caching.}
Storage systems are major target domains of recent in-network computing works (e.g., ~\cite{netcache,netlr,yu20,netstore,netclone,harmonia,pegasus,triaging,switchtx,p4db,switcharoo,tea,concordia,horus,netchain,p4xos}).
\sota~\cite{netcache} shows the potential and limitations of switch-based caching.
The limitation in the item size is passed down to recent works as well.
FarReach~\cite{farreach} enables write-back caching for \sota.
DistCache~\cite{distcache} proposes a distributed caching architecture using power-of-two choices and hash-based cache allocation to balance multiple clusters.
These solutions are based on the \sota architecture, hence they can still cache a limited size of items.
\sys enables variable-length in-network caching by leveraging built-in features of the switch.

\textbf{Other alternatives.}
SwitchKV~\cite{switchkv} employs the Openflow switch for cache lookups while items are cached in a cache node.
Although this reduces the lookup overhead, it still provides limited performance due to server-based caching.
Pegasus~\cite{pegasus} is a selective replication solution where the switch replicates hot items across storage servers for load balancing.
Unfortunately, its performance is limited by the aggregate throughput of storage servers.
\sys provides high-performance single-box caching beyond the aggregate throughput of commodity storage servers. 
\section{Conclusion\label{conclusion}}
We proposed \sys, an in-network caching architecture that is capable of variable-length caching in programmable switches.
The key idea of \sys is to make cached key-value pairs revisit the switch data plane continuously by leveraging packet replication efficiently.
We have implemented a \sys prototype, and experimental results demonstrated that \sys can balance skewed workloads with diverse conditions.
We hope that this work can contribute to the research community by providing insights that utilizing built-in hardware features has great potential to make in-network computing solutions effective.

\label{lastpage1}
\bibliographystyle{plain}
\bibliography{./gykim}
\label{lastpage2}
\end{document}